# Dynamic Multimedia Content Retrieval System in Distributed Environment

R. Sivaraman
Deputy Director
Center for convergence of technologies
Anna University Tiruchirappalli
Tiruchirappalli, India
e-mail: rs@tau.edu.in

R. Prabakaran
Lecturer,
Department of Electrical and Electronics Engineering
Anna University Tiruchirappalli
Tiruchirappalli, India
e-mail: hiprabakaran@gmail.com

S. Sujatha
Lecturer,
Department of Computer Science and Engineering
Anna University Tiruchirappalli
Tiruchirappalli, India
e-mail: ssujtha71@yahoo.co.in

*Abstract*— WiCoM enables remote management of web resources. Our application Mobile reporter is aimed at Journalist, who will be able to capture the events in real-time using their mobile phones and update their web server on the latest event. WiCoM has been developed using J2ME technology on the client-side and PHP on the server–side. The communication between the client and the server is established through GPRS.

Mobile reporter will be able to upload, edit and remove both textual as well as multimedia contents in the server.

## I. INTRODUCTION

A content management system (CMS) is a system used to manage the content of a Web site. Typically, a CMS consists of two elements: the content management application (CMA) and the content delivery application (CDA). The CMA element allows the content manager or author, who may not know Hypertext Markup Language (HTML), to manage the creation, modification, and removal of content from a Web site without needing the expertise of a Webmaster. The CDA element uses and compiles that information to update the Web site. The features of a CMS system vary, but most include Web-based publishing, format management, revision control, and indexing, search, and retrieval.

The Web-based publishing feature allows individuals to use a template or a set of templates approved by the organization, as well as wizards and other tools to create or modify Web content. The format management feature allows documents including legacy electronic documents and scanned paper documents to be formatted into HTML or Portable Document Format (PDF) for the Web site. The revision control feature allows content to be updated to a newer version or restored to a previous version. Revision control also tracks any changes made to files by individuals. An additional feature is indexing, search, and retrieval. A CMS system indexes all data within an organization. Individuals can then search for data using keywords, which the CMS system retrieves.

WiCoM is a wireless application aimed at helping the general administration of cyber contents while being on the move. It is a wireless cyber content management software running on a java enabled mobile device having GPRS connectivity. It finds application in news reporting agency to administer news site in real-time.

A reporter arriving at the site of the event can record the news of the current scenario from the various sources. He can take snaps, audios and videos and upload them right at the moment to the web-server making it available to the world in no time. There are options to edit/delete and thus provide various content management related features. Also, a modified version of it can be useful for e-commerce sites and online shopping sites too.

## II. RELATED WORK

The integrated Content Management System (CMS) is a robust, easy-to-use web content manager built upon a flexible application framework; this framework was developed using inexpensive, open-source resources. It enables users to easily collaborate on creating and maintaining web site content, and provides the contractual relationships between the roles of web site developers, graphic designers, and managers, ensuring quality and integrity of content at all times.

CMS is suitable for just about any web site model, such as news publications, customer support interfaces, Web portals, communities, project management sites, intranets, and extranets. Features include role definitions and workflow customizability, integrated searchable help, a clean modular system for extending the administrative interface, front-end content editing, embedding components into pages, email distribution lists, a news application, discussion forums, and much more. Planned enhancements include content syndication and aggregation, advanced role definitions and workflow customizability and modules. CMS currently requires PHP and Apache. It has been tested on Linux and Windows environments; and while not currently supported, it should run on MacOS X as well. The system natively runs on a MySQL database; however, by using the integrated database abstraction layer, it is possible to use most popular database systems including Oracle, Interbase, MySQL and MS SQL Server.

## III. DEVELOPMENT ENVIRONMENT

The Integrated Development Environment (IDE) chosen was Netbeans 6.0, an open source project that was developed in order to provide a vendor neutral IDE for developing software. This ran on top of the Java SDK1.4_12 which implements functionality that aids in the development of J2ME applications. This functionality includes fully integrating the Java wireless toolkit, which provides configurations and profiles for mobile development. The plug-in also integrates the MIDP2 emulator that is provided with the wireless toolkit which can then be launched from within Netbeans 6.0.

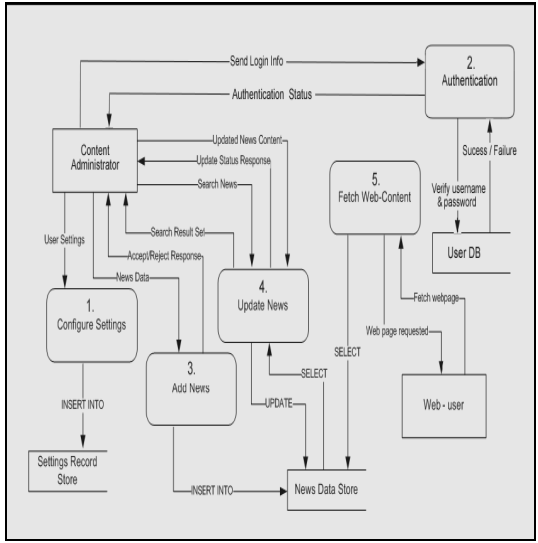

**Figure1: Data flow diagram**

## IV. SYSTEM STRUCTURE

The structure of the system is divided into two components:

- The client-side MIDlet application which resides on the mobile phones,.
- The server-side PHP/MySQL based application.

### 1. Client Side

The client-side system is a MIDlet application which serves as an interface to feed in the contents and control instructions which is interpreted on the server and the appropriate action is taken. The MIDlet has the task of creating textual news contents,

Creating media contents as well as editing and updating textual news contents. The news creation task is done through a data entry interface which contains various sections to be filled. Once done the data is uploaded to the server and stored in the database server.

The media news capture is the most important section of the MIDlet application. It has options to capture pictures, audios as well as videos for the devices that support it. These media can then be uploaded to the server and stored in a particular directory structure. Another most important section of the MIDlet is the News Manager, i.e., the section that helps edit and update news contents posted earlier. It has a search option to search for the relevant news and then bring about a change in it. Once the change is confirmed it is updated onto the database on the server

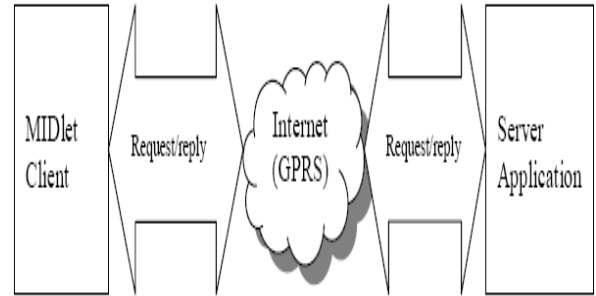

**Figure 2:.Communication between MIDlet and server application**

### Server Side

The server-side system comprises of a web-server, i.e., apache. The scripts are PHP-based while the backend database server is MySQL. The news contents control instructions sent by the MIDlet client is received at the server end and processed by the respective PHP script.

The PHP scripts that handle the MIDlet interaction perform various database based queries and also helps in the generation of xml based data to be consumed by the MIDlet. It is also responsible for dumping of media data properly. The use of xml provides ease in the generation of data for the consumption by the MIDlet. The user-interface is a simple web interface which displays the news contents by fetching them from the server depending on the criteria. The admin-interface is basically for administering the news contents from the desktop.

## V. COMMUNICATION

Wireless content management is a client – server architecture based system, the information flow is not standalone, rather it goes through the network and hence a communication media is needed. The J2ME MIDLets can operate over, and make use of the WAP stack to perform HTTP network interaction, without requiring TCP/IP. Since the server application resides on a remote machine a connection needs to be established between the mobile device and the remote server which can be accomplished with the use of the phones with GPRS connection.

## VI. SYSTEM IMPLEMENTATION

The application begins working from the MIDlet which is the source for input of news content. As the MIDlet is opened a welcome screen is encountered which is followed by a Login Form. Login Form becomes important because of the fact that the system will be used for administration and will require entry into a restricted area of the web-site.

Once the user is authenticated properly the main menu becomes visible and the user can perform the required operations. Once the data is filled completely the upload button can be pressed to bring us to the next screen where the confirmation of data is made. Once done we can send the data to the server by pressing the start button.

The media creation handles the creation of multimedia files, such as, pictures, audios and videos. The camera is initialized by use of various APIs in the MMAPI, e.g., JSR135. Once the file is created it is transferred over the http to the remote server as multipart/form-data.

The News Manager, i.e., the editing and updating section of the application is one of the most important sections. Once the update section is opened the user is provided with a screen where he can enter search keyword as well as the search type such as title, news text or author name. Accordingly the search is performed on the server and all results matching the criteria are fetched. Since the size of the mobile screen is small the data is broken into segments, each containing five types of news in full. This news segment of five is then transmitted to the MIDlet from the server in xml format. In case when more pages exist, the news manager has an extra command option "NEXT" to jump onto the next page else the option is not available.

The use of xml is governed by the kXML parser which is a low footprint xml parser for mobile devices. The client application discussed above was tested on the emulator provided by J2ME wireless toolkit version 2.2 using JAVA SDK 1.4.2_12.

Together with the client-server based system working between the mobile device and the server there is another web-based mobile independent part of this application. This is the website which allows the user to look through various news contents.

There is an admin interface as well for managing few features of the news like activating/deactivating the news from being viewed and also for deleting. Along with all these there is an installation of PHP script, which allows the user to properly configure the server side of the application and set it up properly with ease. The server side of the application is implemented using PHP 5.2 as the language and MySQL 4.1 as the database server with the use of apache 2.0.8 (for windows) as web-server.

## VII. SYSTEM DESIGN AND RESULT ANALYSIS

*1. Server Modules*

1) *User Registration:* User Registration has the screens for registering new users for uploading messages to the server. The registration form gets the information from the user such as first name, last name, user name, password and etc. The new users will informed if they made any errors when they fill the form. If the required information is filled then the new user registration conformed to "registration successful" message.

2) *Message Creator:* Message Creator has the screens for creating the messages on the web server via the web interface. The users can use the different styles like the windows word for creating the text multimedia message. Users can upload their audio and image files using this module.

3) *Message Viewer:* Message Viewer is used to view the uploaded messages on the main page of the Server site. The messages are arranged as descending order according their upload time with the attached image. The message title has the link which leads to the separate page for viewing the full message.

*Client Modules*

1) *Message Creator:* Message Creator is used to create the message using the mobile phone. This module divided into three sub modules namely Text Message Creator, Multimedia Content Creator and Message Uploader. These sub modules described below.

**Text Message Creator** - has the form for getting the Message Title, Content, and Place and Category information. In this form if the specified message categories already exist on the server, then the uploaded message will placed under the specified category on the server web site.

**Multimedia Content Creator** - has the forms for capturing the image using the mobile phone camera and record the audio using the mobile phone microphone. The captured image stored locally on the mobile in the format of jpeg. The recorded audio stored in the format of mp3.

2) *Message Up Loader:* This Message up loader is used to upload the text and the multimedia content to the server. This module has the form which shows the progress of the upload status using the Gauge control. The text message has the higher priority, so it uploaded first, then the multimedia content uploaded finally. This module has the menu option for saving the uploaded message locally with the attachment.

3) *RSS Reader:* RSS Reader module is used to view the contents of the server. The messages on the server are arranged under message categories. This form displays the categories on the mobile screen. The message titles are displayed when the user clicked the category. The full message without the attachment is displayed when the user clicks the message title.

4) *Message Editor:* Message Editor is used to edit the previously stored messages and upload the edited messages to the server. If the already stored the created messages then the new menu item is created with the name of "Saved Items" on the main menu. This menu used to traverse the previously stored messages.

5) *Configurator:* Configurator has the form to get the User Name, Password and Server URL from the server. This form displayed as the first page when the user uses this software for the first time. This information will be modified via the "Edit Data" menu.

is in progress which is a more generalized mobile CMS and eligible for a large scale deployment.

More functions can be added from the prototype design to achieve game content, animation content and movie content. The CMS can also be extended to CDMA technology to support various group of mobile phones. In future, we can think of downloading the content from server to client device for later offline use

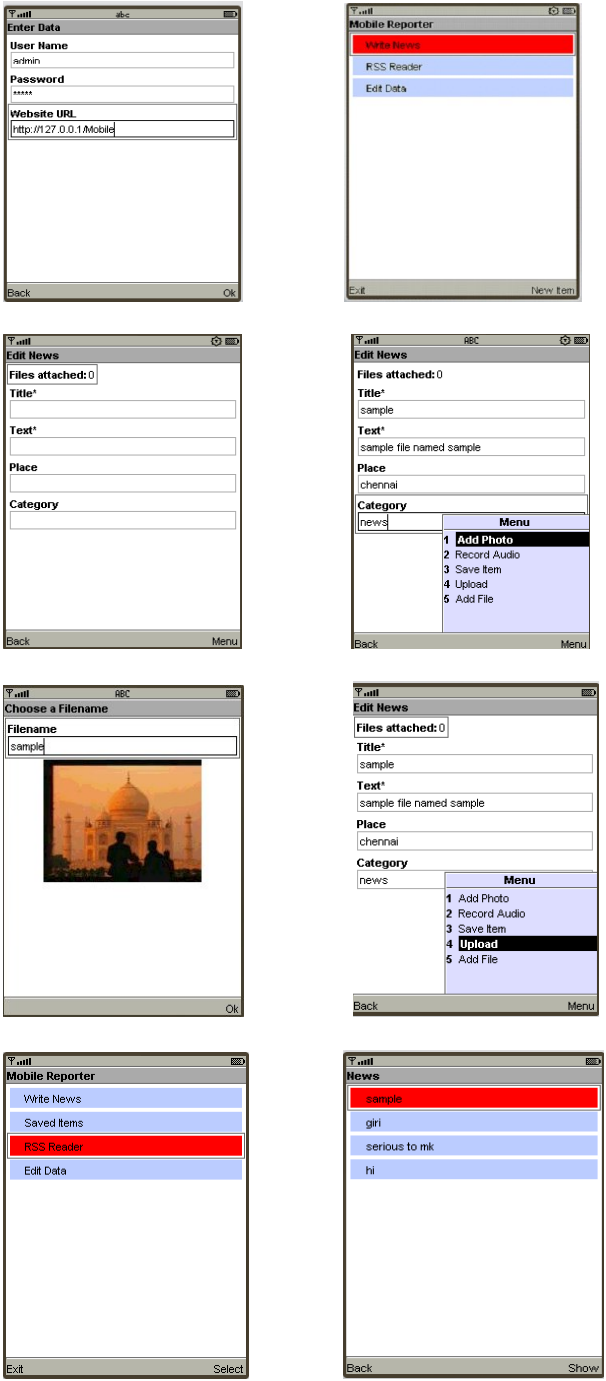

**Figure 3. J2ME Display**

## VIII. CONCLUSION

In this paper we have presented a multimedia application developed using java micro edition, PHP/MySQL on the server side. The use of XML was a good idea keeping in mind the future scope of the project. The application is one of its kinds and finds huge application in news reporting agencies and e-commerce sites. An advanced version of the application